\documentclass[twocolumn,showkeys,showpacs,superscriptaddress,preprintnumbers,amsmath,amssymb,prl]{revtex4-1}
\usepackage{graphicx}
\usepackage{epstopdf}
\usepackage{dcolumn}
\usepackage{bm}
\usepackage{amsmath}
\usepackage{color}
\usepackage{hyperref}
\usepackage{pdfpages}

\usepackage{etoolbox} 

\makeatletter
\patchcmd{\@outputpage@head}{\@ifx{\LS@rot\@undefined}{}{\LS@rot}}{}{}{}
\makeatother

\begin{document}

\title{`Sinking' in a bed of grains activated by shearing}

\author{Hu Zheng}
\email{hzheng@phy.duke.edu}
\affiliation{School of Earth Science and Engineering, Hohai University, Nanjing, Jiangsu,China}
\affiliation{Department of Physics \& Center for Nonlinear and Complex Systems, Duke University, Durham, North Carolina 27708, USA}
\author{Dong Wang}
\affiliation{Department of Physics \& Center for Nonlinear and Complex Systems, Duke University, Durham, North Carolina 27708, USA}
\author{Jonathan Bar\'{e}s}
\email{jb@jonathan-bares.eu}
\affiliation{Department of Physics \& Center for Nonlinear and Complex Systems, Duke University, Durham, North Carolina 27708, USA}
\affiliation{LMGC, UMR 5508 CNRS-University Montpellier, 34095 Montpellier, France}
\author{Robert P. Behringer}
\affiliation{Department of Physics \& Center for Nonlinear and Complex Systems, Duke University, Durham, North Carolina 27708, USA}

\begin{abstract}
We show how a weak force, $f$, enables intruder motion through dense granular materials subject to external mechanical excitations, in the present case stepwise shearing. A force acts on a Teflon$^{\mbox{\scriptsize{\textregistered}}}$ disc in a two dimensional system of photoelastic discs. This force is much smaller than the smallest force needed to move the disc without any external excitation. In a cycle, material + intruder are sheared quasi-statically from $\gamma = 0$ to $\gamma_{max}$, and then backwards to $\gamma = 0$. During various cycle phases, fragile and jammed states form. Net intruder motion, $\delta$, occurs during fragile periods generated by shear reversals. $\delta$ per cycle, \textit{e.g.} the quasistatic rate $c$, is constant, linearly dependent on $\gamma_{max}$ and $f$. It vanishes as, $c \propto (\phi_c - \phi)^a$, with $a \simeq 3$ and $\phi_c \simeq \phi_J$, reflecting the stiffening of granular systems under shear\cite{ren2013_prl} as $\phi \rightarrow \phi_J$. The intruder motion induces large scale grain circulation. In the intruder frame, this motion is a granular analogue to fluid flow past a cylinder, where $f$ is the drag force exerted by the flow.
\end{abstract}
\date{\today}


\maketitle

How can an intruder, subject to a weak force, burrow through a dense granular material undergoing mechanical excitation? Such motion has drawn considerable recent interest \cite{kolb2004_pre,kolb2013_pre,seguin2011_prl,harich2011_epl,staron2015_pre}, and is relevant to fields from biology \cite{maladen2009_sci}, and earth science \cite{clark2012_prl,clark2013_epl,clark2014_pre} to engineering. Intruder motion implies temporary weakening of the material due to mechanical excitation, which may be vibration or tapping. Here, however, we consider a different excitation: quasi-static cyclic shear.
Particularly relevant is the granular jamming/unjamming transition between fluid-like and solid-like states \cite{hecke2010_jpcm,bi2011_nat,reddy2011_prl,zhang2010_gm,hartley2003_nat,ren2013_prl}, where the packing fraction, $\phi$, is key. Increasing $\phi$ above $\phi_J$ jams the system \cite{liu1998_nat,hecke2010_jpcm}, limiting intruder mobility. Shear also controls jamming. For $\phi_S < \phi < \phi_J$, sheared low stress states strengthen and can jam \cite{bi2011_nat,ren2013_prl,majmudar2005_nat}. Reversing shear weakens force networks, unjams fragile \cite{cates98_prl} or shear jammed states, and enhances intruder mobility. Transitioning between unjammed, fragile and jammed states, without changing density, creates long range force changes and plastic rearrangements, facilitating or limiting intruder motion. Overall, mechanical fluctuations are a source of thermal-like noise \cite{reddy2011_prl,debregeas2001_prl,sollich1997_prl,Schall2007_sci,behringer2008_prl,kolb2004_pre,harich2011_epl}. 

\begin{figure}[!b]
\includegraphics[width=7cm]{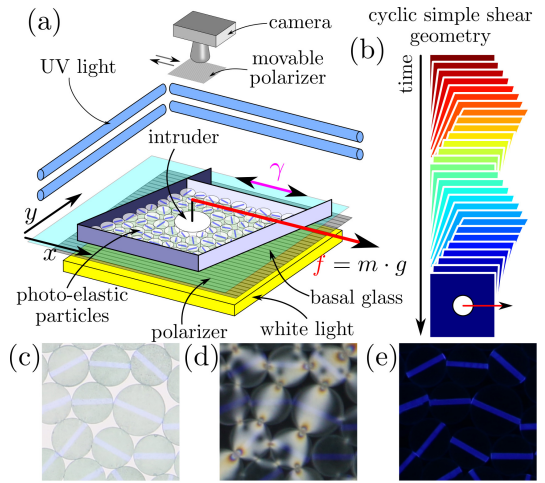}
\caption{\label{fig:setup} (color online). (a) Sketch of experimental setup: A $76.2$mm diameter intruder immersed in a bidisperse layer of photoelastic disks was pulled with a constant force $f$. The intruder started at low $x$ values (\textit{i.e.} to the left), centered along the $y$-axis; under shear, it moved along the $x$-axis. The system was imaged from above by a high-definition camera following each shear step:  in white light (c), cross-polarized (d) light and UV (e). Each particle was marked with a small UV-sensitive bar. (b) Shear protocol: One cycle of shear strain was applied to the whole system stepwise to $\gamma_{max}$ by deforming the initially square box to a parallelogram, at constant area. The shear was then reversed, also stepwise, back to the original boundary configuration. This protocol was repeated $N_c$ times.}
\end{figure}

We characterize the motion of a large (relative to grains) embedded object, subject to time-varying cyclic shear, and a force, $f$, parallel to the shear direction. These experiments provide two key new features: ($i$) control of overall density; ($ii$) observation of particle scale forces and motion. These provide for the first time to our knowledge, direct connections between granular mechanics and intruder motion. If $f = 0$, the intruder executes random motion. Otherwise, even if $f$ is much smaller than the smallest force needed to drive the intruder through the unperturbed material, it advances slowly in the $\vec{f}$ direction. The intruder experiences granular stresses that are low when the material is unjammed, fragile, or very weakly jammed \cite{geng2005_pre,candelier2009_prl,candelier2010_pre,bi2011_nat,coulais2014_prl}. Crucially `unjamming' during cyclic shear provides deformable granular states enabling much of the intruder motion. Force chains/networks \cite{bi2011_nat,zheng2014_epl} play a key role, changing dramatically over a cycle. Shear/shear-reversal jams/weakens granular systems \cite{bi2011_nat,ren2013_prl,zheng2014_epl}, which switches the intruder mobility from low to high. As the intruder moves through the material, grains flow around it. In the intruder frame, this system is a granular analogue of fluid flow around a cylinder. By Newton's third law, $f$ is the drag force exerted by the granular material on the intruder, and the locally coarse-grained flow shows vortices resembling fluid turbulence.

Previous experiments explored related phenomena: studies of intruder speed vs. drag force \cite{takehara2014_prl}, granular flow around an intruder \cite{kolb2004_pre,seguin2011_prl,harich2011_epl,seguin2013_pre,kolb2013_pre}, and slow creep of an intruder outside of a shear band \cite{reddy2011_prl}. However, a deep understanding of cooperative microscopic granular dynamics leading to the object's motion, addressed here, does not exist to our knowledge \cite{goyon2008_nat}.

Our intruder is a $7.62$cm diameter Teflon$^{\mbox{\scriptsize{\textregistered}}}$ disc embedded in a quasi-statically sheared 2D granular medium of bidisperse photoelastic discs, subject to a uniform and constant drag force, $f$, along the shear direction. Intruder and grains are confined in a horizontal cell \cite{zheng2014_epl}, Fig.\ref{fig:setup}-(a), and cyclically sheared $N_c$ times, with amplitude $\gamma_{max}$. Shear is carried out stepwise by deforming initially square boundaries ($40 \times 40$cm$^2$) to a parallelogram, then returning to their original shape. Each step has magnitude $0.25\%$  (Fig.\ref{fig:setup}-(b)), with $1.25\% \le \gamma_{max} \le 20\%$. $N_c = 50$, unless the intruder reaches the boundary. Photoelastic particles provide grain-scale forces. The bi-disperse discs have diameters $1.59$cm and $1.27$cm and thickness $0.64$cm, were cast from photoelastic material, as in \cite{cox2016_epl,bares2017_epj}, and have small UV-sensitive lines on top. The inter-grain friction coefficient was $0.62$. The packing fraction $\phi$ was constant for a given run, and was tuned by changing the particle number from 852 to 896, with the ratio of large to small particles constant ($N_{large}/N_{small}=0.29$). A constant $f$ was applied by a string running just above the particles and over a pulley, then to a vertically hanging mass, $m$ ($64g \le m \le 109 g$). $f$ was too small to move the intruder without shear (see supplementary material). The system was illuminated from below by circularly polarized white light, and from above by UV light. After each shear step, we recorded high resolution images (Canon EOS 70D, $5472 \times 3648$ px$^2$) with and without a crossed circular polarizer and with only UV light on (see Fig.\ref{fig:setup}-(c), (d) \& (e) respectively). Normal light/UV light images gave particle and intruder positions/orientation. The pressure, $P$ inside each disc (Fig.\ref{fig:AverageLocal}-(c) to (f)) was computed from the squared gradient of the photo-elastic image intensity as in \cite{howell_prl1999,geng_prl2001,ren2013_prl,behringer_jsm2014}; $P \propto \overline{G^2}= \langle |\nabla I|^2 \rangle$ \cite{majmudar2005_nat,geng2005_pre} where the bar implies a system average. 

\begin{figure}[!b]
\includegraphics[width=6.5cm]{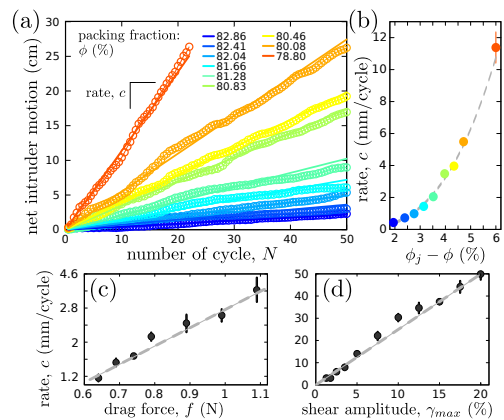}
\caption{\label{fig:GlobalBehaviour} (color online). (a) Cumulative net intruder motion $\delta$ \textit{vs.} $N$, for different constant for different packing fractions $\phi$. Here, $f=1.09$N and $\gamma_{max}=10\%$. The data are consistent with a linear dependence of $\delta$ on $N$: $\delta = c \cdot N$. (b) Variation of the quasistatic rate $c$ as a function of the distance to the jamming packing fraction. This varies as a power-law: $c=0.054 (\phi_J - \phi)^{2.95}$ ($\phi$ in percent). (c) Constant drag force $f$ \textit{vs.} $c$. Here, $\phi= 80.83\%$, and $\gamma_{max}=10\%$. The intruder quasistatic rate increases linearly with $f$: $c = 4.28 \cdot f - 1.46$~mm. (d) Shear amplitude $\gamma_{max}$ \textit{vs.} $c$ with $f=1.09$N and $\phi= 80.83\%$. $c$ increases linearly with $\gamma_{max}$: $c = 2.56 \cdot \gamma_{max} + 1.11$~mm.}
\end{figure}

{\bf Characterizing the response:} The intruder exhibits affine and non-affine motion. In the $y$-direction, the intruder, which sits approximately in the middle, experiences only small-amplitude, roughly periodic, zero-mean non-affine motion (although we observe some random drift--see Fig.S1). In the $x$-direction, it experiences affine motion due to forward shear of $\sim \gamma_{max}/2$, and a nominal return to its starting point during reversal (Fig.\ref{fig:setup}-(b)), hence zero affine motion per cycle.  But, the non-affine intruder motion, $\delta_x$, parallel to the applied force, provides net forward motion each cycle. Henceforth, we only consider $\delta \equiv \delta_x$. Several parameters affect $\delta$: $\textit{f}$, $\phi$, $\gamma_{max}$, the inter-granular friction coefficient, the intruder shape and relative size, the shear box dimension, \textit{etc}. Here, we focus on the first three. 

We first consider transients and boundary effects. For $5$ to $10$ cycles, the grains rearrange, forgetting their initial conditions. Also, $\delta$ varies more slowly for higher $N$ as the intruder approaches the boundary, an effect that is stronger for larger $f$, as the intruder comes closer to the boundary then. Except for a few initial and final cycles, $\delta$ varies linearly with $N$: $\delta = c \cdot N$. This behavior, Fig.\ref{fig:GlobalBehaviour}(a), occurs for all$\phi$, $f$, and $\gamma_{max}$ (see Supplementary Material). 

The key global scale physics is contained in $c(f,\phi,\gamma_{max})$. To characterize the role of packing fraction we vary $\phi$ from $78.8\%$ to $82.86\%$ for fixed $\textit{f}=1.09$N and $\gamma_{max}=10\%$. The lowest packing fraction is slightly above the lower limit of shear jamming, and the highest packing fraction is a bit lower than the isotropic frictionless jamming point \cite{bi2011_nat,ren2013_prl,zheng2014_epl}. For these $\phi$'s, $\delta$ is linear in $N$, Fig.\ref{fig:GlobalBehaviour}-(a), but as in Fig.\ref{fig:GlobalBehaviour}-(b) $c$ follows a power-law in the distance to $\phi_J \simeq83.5\%$: $c \sim (\phi_J-\phi)^n$, with $n = 2.95 \pm 0.05$. Thus, for higher $\phi$'s, the intruder experiences diverging resistance from the granular material. This is consistent with data of Ren et al. \cite{ren2013_prl}, who showed a diverging Reynolds pressure, {\it i.e.} pressure generated by shear strain, characterized by the Reynolds coefficient: $R \sim (\phi_J - \phi)^{-\beta}$, where $\beta = 3.3 \pm 0.1$ ($R \equiv (1/2) \partial^2 P/\partial \gamma ^2$). We probe the role of $f$ by varying $f$ from $0.64$N to $1.09$N at fixed $\phi = 80.83\%$, and $\gamma_{max} = 10\%$. As shown in Fig.\ref{fig:GlobalBehaviour}-(c), $c$ varies linearly with $f$, $dc/df = 4.28 \pm 0.34 mm/N$ \footnote{Error bars give the $95\%$ confidence interval.}, unlike the `Eyring-like' \cite{eyring1936_jcp} behavior observed by \cite{reddy2011_prl}. $c$ depends linearly on $\gamma_{max}$, e.g. Fig.\ref{fig:GlobalBehaviour}-(d) giving the variation of $c$ for $\gamma_{max} = 1.25 \%$ to  $\gamma_{max} = 20 \%$: $\gamma_{max}$; $d c/d \gamma_{max} = 2.56 \pm 0.15 mm$.

\begin{figure}[!t]
\center
\includegraphics[width=8.5cm]{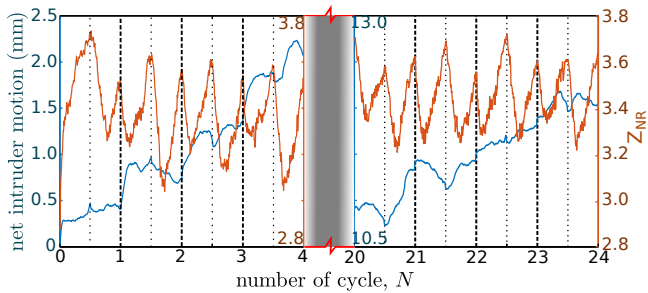} 
\caption{\label{fig:Z-v-step} (color online) Net intruder displacement (blue) and average contact number for non-rattler particles (red) of the first four cycles and $20^{th}$ to $24^{th}$ cycles ($f = 1.09$N, $\phi = 80.83\%$ and $\gamma_{max} = 20\%$).}
\end{figure}

A related controlling factor is whether the system is jammed, fragile or unjammed over different parts of each cycle. The contact number, $Z$ is a useful measure of jamming, and Fig.~\ref{fig:Z-v-step} shows data for the system-averaged $Z$ for non-rattler particles, {\it i.e.} particles with at least two contacts. Zhang et al. \cite{bi2011_nat} showed that shear jamming for a similar system was reached for $Z_{NR} \simeq 3.4$, with fragile states down to $Z_{NR} \simeq 3.0$. Fig.~\ref{fig:Z-v-step} makes several points. First, $Z_{NR}$ varies between values fragile state values up to robustly jammed values where $Z_{NR} \simeq 3.7$. Under shear reversal, $Z_{NR}$ \textit{vs.} strain (in step number) changes slope abruptly. Typically, the intruder experiences much of its forward motion between the strain direction reversal and when $Z_{NR}$ reaches a small value. Strain reversals cause changes in the orientation of the force network, and the minima in $Z_{NR}$ correspond to times when the system is softest. 

The softening and reorientation of the network facilitate the intruder motion. As noted, $\overline{G^2}$ is roughly linearly proportional to the pressure \cite{majmudar2005_nat,geng2005_pre}. During a cycle, the difference in forces acting on the left and right sides of the intruder changes substantially, becoming small or zero following strain reversals, and in general when $\delta > 0$. Also, the material stiffens and softens over a cycle, limiting or enhancing mobility. Additionally, there is asymmetry in the granular stresses between the left and right side of the intruder. Finally, shear reversals reorient the force networks, and hence the principal stress directions, causing wiggling of the intruder and changes in the force on the intruder in the $x$-direction. Below, we elaborate on these features.

\begin{figure}[!t]
\center
\includegraphics[width=8.5cm]{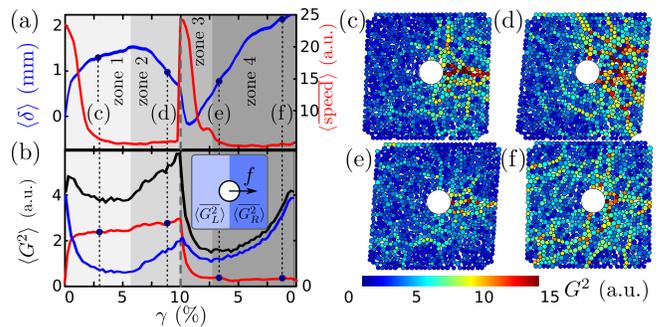}
\caption{\label{fig:AverageLocal} (color online). (a) Net intruder displacement (blue) and mean particle displacement (red) during a typical cycle ($f = 1.09$N, $\phi = 80.83\%$ and $\gamma_{max} = 10\%$). (b) Evolution of $P$ to the right of the intruder, $\langle \overline{G^2_R} \rangle$ (black), to the left of the intruder, $\langle \overline{G^2_L} \rangle$ (blue), and the difference: $\langle \overline{G^2_R} \rangle - \langle \overline{G^2_L} \rangle$ (red). The last measures the net granular drag force on the intruder. Inset: sketch of the regions to the Left (light blue) and Right (dark blue) of the intruder center, used to average $G^2$. (c), (d), (e) and (f): $G^2$ for each particle for different steps of the shear cycle presented in (a) and (b). In (c), the intruder compresses grains to the right. In (d), the shear jammed state is nucleating. The intruder moves in the negative $x$-direction because of high differential granular pressure. In (e), following shear reversal, the particles rearrange and the granular force network almost vanishes. In (f) the force network has reformed but it is weaker than and has a different orientation than the network in (d).}
\end{figure}

In the first cycle, the system starts from $P = 0$. During the first forward shear, the network and $\overline{G^2}$ initially grow rapidly, then more moderately and typically saturate as $\gamma \rightarrow \gamma_{max}$. When the strain is reversed, a similar process occurs, but the state at reversal, set by the forward shear, is strongly jammed, with a network that more robustly resists forward than reverse shear. After a few cycles, the system behavior is typified by the cycle shown in Fig.~\ref{fig:AverageLocal} (see videos in SM).
The initial state of this cycle inherits a force network created by the previous reverse strain. 

To understand the effect of the force networks on the intruder motion, we separate the granular system into Right (in `front' of the intruder) and Left (behind the intruder--see inset of Fig.\ref{fig:AverageLocal}-(b)) along a vertical line through the intruder center. The difference $\langle \overline{G^2_R} \rangle - \langle \overline{G^2_L} \rangle$, (e.g. Fig.\ref{fig:AverageLocal}-(a)) is a good measure of the net granular force on the intruder, which we relate to the net intruder motion in Fig.\ref{fig:AverageLocal}-(b). Fig.\ref{fig:AverageLocal} pertains to a typical cycle after transients with $\gamma_{max}=10\%$, $\phi=80.83\%$, and $f=1.09$N. We also relate the measures of Fig.\ref{fig:AverageLocal}-(a,b) to representative force networks during several phases of the cycle, Fig.\ref{fig:AverageLocal}-(c-f). For forward and reverse shear, the response immediately after shear reversals shows strong variations in $\langle \overline{G^2_R} \rangle - \langle \overline{G^2_L} \rangle$, in the force networks, and the displacements of the intruder and the grains. In particular, $\langle \overline{G^2_R} \rangle - \langle \overline{G^2_L} \rangle$ decreases strongly after reversals as the networks, which have oriented to resist the applied shear, weaken and reorient to resist shear in the reverse direction. The states following reversals are relatively fragile/soft, allowing $\delta$ to grow. This unloading mechanism, in the present case, associated with shear unjamming, is the crucial mechanism that allows the intruder to advance over multiple cycles. After some amount of strain following a reversal, the system re-jams, limiting the intruder non-affine motion, and setting the stage for the next shear reversal. Stress reversals are associated with oscillatory motion of the intruder in the $y$-direction, but with almost no net vertical motion over a cycle.

\begin{figure}[!t]
\includegraphics[width=8cm]{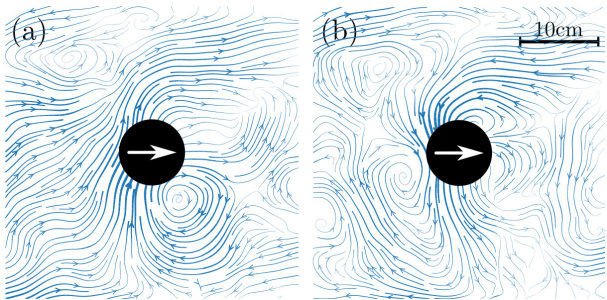}
\caption{\label{fig:flow} (color online). Large scale quasistatic convective flow. Images (a) and (b) show the non-affine flow of particles in the frame of the intruder averaged over $50$ cycles corresponding to positions (c) and (e) of the shear cycle in Fig.\ref{fig:AverageLocal}-(a)-(b) respectively. There is large scale collective motion extending to the boundaries of the box including major vortices that partially change direction between (a) and (b). The average non-affine displacement of the particles is given in Fig. S2 of the SM. These data pertain to a typical experiment: $f = 1.09$N, $\phi = 80.83\%$ and $\gamma_{max} = 15\%$. See SM for a video of the quasistatic flow.}
\end{figure}

{\bf Collective particle motion:} The intruder motion leads to system spanning collective  particle motion that varies with the phase. In the intruder frame, this motion is reminiscent of fluid flow past a cylinder. Parts (a) and (b) of Fig.\ref{fig:flow} show the cycle-averaged particle displacement streamlines in the intruder frame at times marked by c and e in Fig.\ref{fig:AverageLocal}-(a). Each corresponds to about 1/8 of a cycle after respectively, the reverse-forward, and forward-reverse strain direction changes.  Overall, the material must flow from in front of the intruder to behind it, and the overall particle displacements/cycle, averaged over all cycles of a run are given in Fig. S2 of Supplementary Material. From the lab frame perspective, these collective motions occur as the intruder pushes material to the right/left respectively in Fig.\ref{fig:flow} (a)/(b), and are not time-reversal symmetric for the two cases. Particles undergo quasistatic flow everywhere in the cell and their motion is largest when the intruder quasistatic rate is largest. (see SM for details and a movie).

We contrast the present large scale grain motion with the results of Kolb et al. \cite{kolb2004_pre,kolb2013_pre} and of Harich et al. \cite{harich2011_epl}. The Kolb et al. \cite{kolb2004_pre} flows consist roughly of two counter-rotating vortices, that are qualitatively but not necessarily quantitatively similar to the collective motion observed in Fig.S2. These authors did not report the complex turbulent-like flow seen here. More recently, Kolb et al. \cite{kolb2013_pre} observed flow throughout their system, with a return flow behind the intruder, hence, no large empty wake behind the intruder. The Harich et al.  \cite{harich2011_epl} experiments also show grain flow in the vicinity of the intruder, but seemingly much weaker than in the present experiments. 

To conclude, the present experiments have demonstrated the key physical mechanisms that enable intruder motion through a dense granular material activated by cyclic shear. An essential feature that has not been available in previous studies of intruder motion is the ability to measure particle scale forces. This approach shows that the intruder moves when the force network weakens (following strain reversals), the net granular force on the intruder is small, and the contact number, $Z_{NR}$ drops. The intruder moves through the material at a quasistatic rate, $c$, that varies linearly with the external driving force, $f$, and with the amplitude of shear strain, $\gamma_{max}$. $c$ depends strongly on $\phi$, effectively vanishing at a critical $\phi$ close to the isotropic jamming $\phi_J$. The dependence of $c$ on $\phi$, $c \propto (\phi_J - \phi)^{2.95}$, matches (inversely) the Reynolds pressure dependence on $\phi$ reported by Ren et al. \cite{ren2013_prl}. In the intruder frame, $f$ corresponds to the drag force exerted by grains that flow past the intruder. The grain flow exhibits complex space-filling vortex motion that is qualitatively similar to fluid flow past a cylinder. This similarity is striking, since the fluid case is associated with instability in continuum equations of motion, whereas the grain case corresponds to the collective motion of manifestly discrete particles. The mechanisms that allow intruder motion are not limited to two dimensions, and it seems reasonable to expect that they would also apply in three dimensions.

By comparison, in the work of Reddy et al. \cite{reddy2011_prl} the intruder advances in a direction that is at least partially resisted by the shear stresses induced by the applied shear. The Eyring-like excitations of the intruder in the Reddy et al. experiments suggest that the intruder motion was caused by intermittent fluctuations associated with the steady shear, that provide the mechanism to temporarily reduce the granular force resisting the forward motion of the intruder. Also in Reddy et al. \cite{reddy2011_prl}, grains can rearrange because the surface of the (3D) material is unconstrained.  

Acknowledgements: This work was supported by NSF Grant No. DMR1206351, NASA Grant No. NNX15AD38G, the William M. Keck Foundation, a RT-MRSEC fellowship and Labex NumEv anr-10-labx-20 (JB). HZ also thanks NSFC Grant NO. 41672256 and the NSF (Jiangsu, China) Grant No. BK20140845 for financial support.

\bibliography{biblio}

\newpage


\section*{Supplementary material}

# Supplementary materials for "'Sinking' in a bed of grains activated by shearing"

Hu Zheng, Dong Wang, Jonathan Barés, Robert P. Behringer

December 11, 2017

The intruder motion consists of intruder displacements that are both perpendicular ($y$-direction) and parallel ($x$-direction) to the applied force. As presented in Fig.S1, the perpendicular motion has a nearly zero mean. The motion parallel to $f$ and the shear strain increases linearly with the cumulated shear. **We unfold the cumulated shear by presenting it in terms of the number of (small) steps taken.**

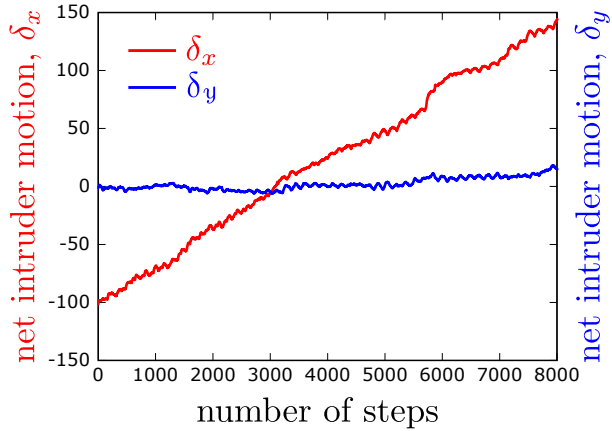

Figure S1: $x$ and $y$ intruder motion for an experiment with drag force $1.09N$, packing fraction $80.83\%$ and maximum shear amplitude $20\%$.

Fig.S2 presents the evolution of the net intruder motion $\delta$ as a function of the number of cycles $N$ when the drag force, $f$, and the maximum shear amplitude $\gamma_{max}$ vary. These curves show the linear variation of $\delta$ with $N$ and correspond to the quasistatic rate $c$ presented in Fig.2 of the main text.

S1

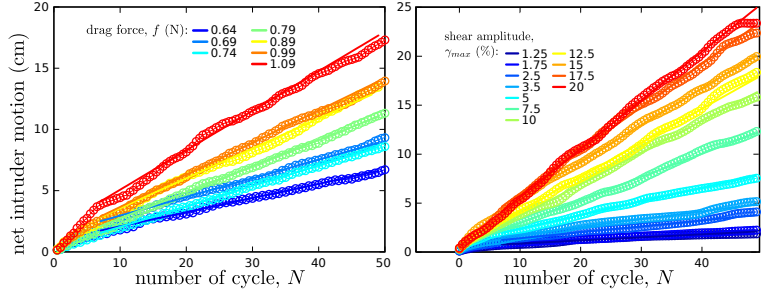

Figure S2: Cumulated net intruder motion $\delta$ vs. the number of shear cycles $N$, for different drag forces, $f$. The packing fraction is $\phi = 80.83\%$, and the shear amplitude is $\gamma_{max} = 10\%$. Linear fits shows that $\delta$ varies linearly with $N$: $\delta = c \cdot N$. $\delta$ vs. $N$ for different packing fraction $\phi$. The drag force is $f = 1.09N$ while the shear amplitude is $\gamma_{max} = 10\%$.

---

Fig.S3 shows the overall particle displacements/cycle, which is averaged over all cycles of a run with: $f = 1.09N$, $\phi = 80.83\%$ and $\gamma_{max} = 15\%$. There is particle flow throughout the cell. Also of note: the intruder speed is maximum when the average particle speed is also maximum.

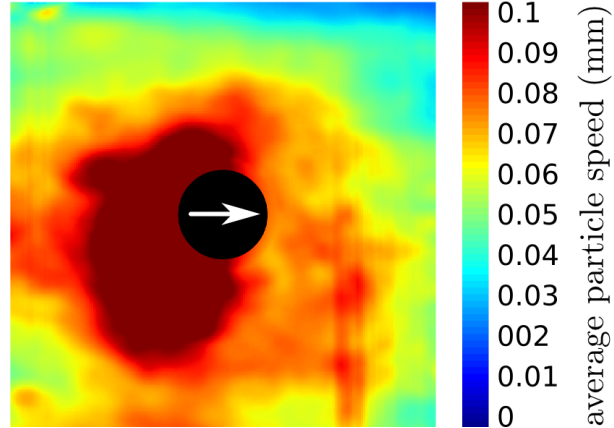

Figure S3: Average non-affine displacement of the particles in a frame centered on the intruder. The displacement field is averaged over all steps of all cycles. These data are extracted from a typical experiment: $f = 1.09N$, $\phi = 80.83\%$ and $\gamma_{max} = 15\%$.

---

'Weak force means that the force experienced by the intruder is not large enough to make it move inside the granular system **in the absence of external excitation**. In Fig.S4, we present measurements where the intruder have been pulled through the granular bed without shearing it. The critical force to move the intruder for different packing fraction is shown in this figure. Even with a small packing fraction $\phi = 78.80\%$, the force needed to move the intruder is at least around 6N which is about 6 times the largest force we applied in our shearing experiment.

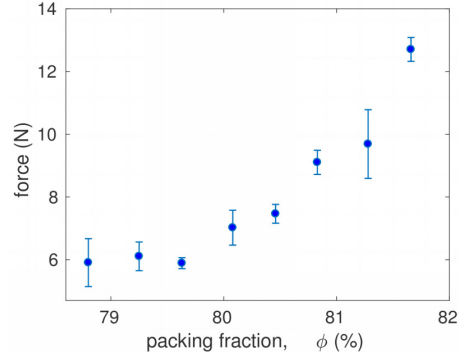

Figure S4: Minimum force to pull the intruder through the granular bed without shearing, as a function of the packing fraction.

---

Fig.S5, we present the probability of big/small particles to have big or small particles as neighbors during the course of an experiment. The probabilities fluctuate around a constant value, which indicates that systematic local segregation does not occur.

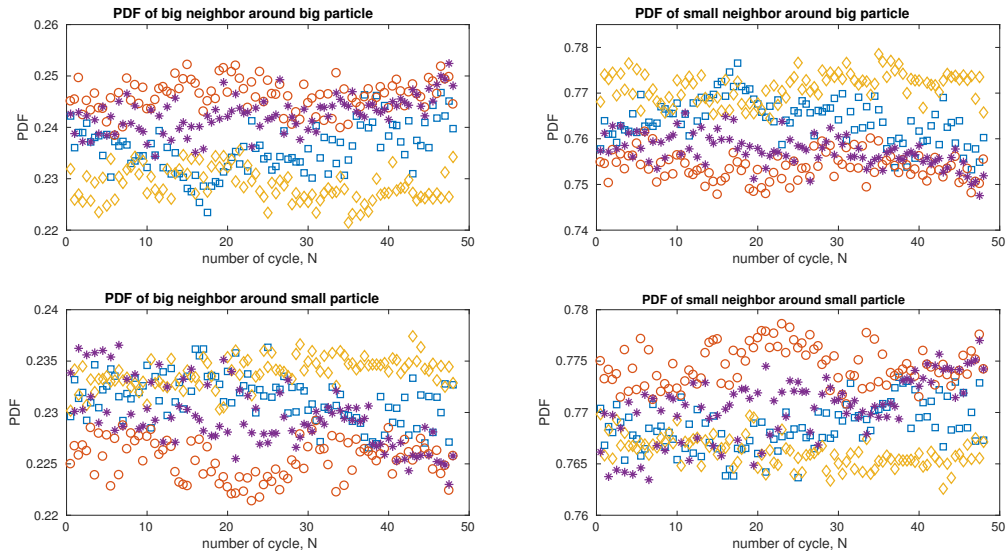

Figure S5: The probability for each particle type (e.g. larger or smaller) to have a given number of neighbors. Different colors indicate different experiments.

---

**Movie1:**
https://youtu.be/n_qkRqFL2ZY

Movie of the evolution of the averaged particle non-affine displacement field during the back and forth shear cycle of a typical experiment (drag force $1.09N$, packing fraction $80.83\%$ and maximum shear amplitude $10\%$). The displacement field is averaged over all the steps of all the cycles.

---

**Movie2:**

<https://youtu.be/gQXaqHjT780>

Movie of the displacement of the particles and intruder during several cycles for a typical experiment (drag force $1.09N$, packing fraction $80.83\%$ and maximum shear amplitude $10\%$). The small bars attached to the particles give their displacement direction; the color, from blue to red, gives the magnitude of the displacement.

---

**Movie3:**

<https://youtu.be/KXJG-vJi_WM>

Movie of the particle rotation when the system is sheared back and forth during several cycles, for a typical experiment (drag force $1.09N$, packing fraction $80.83\%$ and maximum shear amplitude $10\%$). The intensity of the particle color (from light to dark blue) gives the intensity of the rotation of each particle from one shear step to another.

---

**Movie4:**

<https://youtu.be/kwWlH6fWVho>

Movie of the particle pressure when the system is sheared back and forth during several cycles for a typical experiment (drag force $1.09N$, packing fraction $80.83\%$ and maximum shear amplitude $10\%$). The intensity of the particle color (from light to dark blue) gives the pressure for each particle. Force chains appear when the system is sheared.



\end{document}